\documentclass[aps,pre,twocolumn,superscriptaddress,nofootinbib]{revtex4-2}
\usepackage{amsmath,amssymb,bm}
\usepackage{graphicx}
\usepackage{xcolor}
\usepackage{tikz}
\usetikzlibrary{arrows.meta,decorations.pathreplacing}
\usepackage[colorlinks=true,linkcolor=blue,citecolor=blue,urlcolor=blue]{hyperref}

\renewcommand{\P}{\mathbb{P}}

\newcommand{\nul}{\varnothing}
\newcommand{\CE}{\mathrm{CE}}
\newcommand{\M}{\mathsf{M}}

\definecolor{cplus}{HTML}{2A78D6}
\definecolor{cminus}{HTML}{EB6834}
\definecolor{cmask}{HTML}{D9D9D6}

\newcommand{\be}{\begin{eqnarray}} 
\newcommand{\ee}{\end{eqnarray}}
\newcommand{\D}{\mathrm{d}}
\renewcommand{\vec}{\mathbf}
\newcommand{\bsig}{\boldsymbol\sigma}

\begin{document}

\title{Masked diffusion as random sequential adsorption: an exactly solvable model for parallel decoding}

\author{Adrian Baule}
\affiliation{School of Mathematical Sciences, Queen Mary University of London, London E1 4NS, UK}

\date{\today}

\begin{abstract}
Masked diffusion models generate text by reversing a stochastic masking process. Unlike autoregressive LLMs, which generate text one token at a time, they can reveal many tokens in parallel, at the price of an error, since tokens revealed together are drawn independently. We show that this trade-off maps onto the statistical physics of driven many-particle systems. (i) For any data, the error of parallel generation is the entropy production of the generation process relative to the reversed masking process. (ii) For Markov data, generation is a random sequential adsorption process with labelled gaps, which we solve exactly for any number of steps. (iii) For many steps, the error becomes the dissipation of a slowly driven system: the unmasking schedule plays the role of the driving protocol, the number of steps that of its duration, and the error is set by a thermodynamic metric that we obtain in closed form from the mutual information between tokens. Optimal schedules are geodesics of this metric. The exact solution shows how correlations in the text shape both the error and the quality of the output. The longer the correlations, the more the error concentrates at the start of generation. Schedules that start with small steps then gain more, both in path error and in perplexity, and need fewer steps to reach their asymptotic behaviour.
\end{abstract}

\maketitle

Generative models for text are usually \emph{autoregressive}. Text is split into tokens (words, word fragments or characters), and a neural network is trained to predict the probability distribution of the next token given all previous ones. A sentence is then produced one token at a time, from left to right, so producing $L$ tokens takes $L$ sequential evaluations of a large network. Masked diffusion models (MDMs)~\cite{Austin2021,Lou2024,Sahoo2024,Shi2024,Ou2025}, which have recently been scaled to large language models~\cite{Nie2025,Mercury2025}, take a different route. They are trained on a fill-in task: tokens of a text are replaced at random by a blank ``mask'' symbol, and the network learns to predict the hidden tokens from the visible ones. Generation runs this process backwards: it starts from a sequence of masks, and a single network evaluation returns, for \emph{every} masked position at once, the probability distribution of its token given the tokens revealed so far. Many positions can therefore be revealed per evaluation, and a sequence is generated in $N\ll L$ steps.

The price of parallel generation is an error. Tokens revealed in the same step are drawn independently, each from its own distribution, although in the data they are correlated. Each step thus makes a mean-field approximation: the joint distribution of the tokens revealed together is replaced by the product of their single-site marginals. How the resulting \emph{parallel decoding} error depends on the number of steps $N$, on the \emph{schedule} that fixes how many tokens are revealed at each step, and on the statistics of the data is the central trade-off of the field~\cite{Kang2026,LavenantZanella2025,ChenCongLi2026,Wainwright2026}.

Several recent works quantify this trade-off for general data. The error of a single parallel step is the conditional total correlation of the tokens revealed together~\cite{Kang2026,Zhang2026}, and the accumulated error is bounded by the mutual information between tokens~\cite{LiCai2025}. The expected relative entropy for an arbitrary schedule has been expressed through an information profile of the data, from which near-optimal schedules follow~\cite{LavenantZanella2025,ChenCongLi2026}. In the limit of many steps the optimal error is the squared integral of the square root of a complexity density along the reveal path~\cite{Wainwright2026}, a structure familiar from thermodynamic length~\cite{SalamonBerry1983,Crooks2007}. Schedules have also been optimised numerically by minimising a KL upper bound~\cite{Park2025}. Tractable data models used so far include $n$-gram and hidden Markov languages~\cite{Feng2025} and random walks on graphs~\cite{Bansal2026}, and speed--accuracy relations from non-equilibrium thermodynamics have been derived for continuous diffusion models~\cite{Ikeda2025}. What is still missing is a model in which both the error and the quality of the output can be computed exactly at any number of steps, so that the large-$N$ theory can be tested directly.

In this paper we provide such a model. For data generated by a Markov chain over an arbitrary vocabulary, the bigram language model~\cite{Shannon1948}, generation maps onto a random sequential adsorption (RSA) process~\cite{Renyi1958,Evans1993,Baule2017,Baule2019}: revealed tokens behave like deposited particles that shield the gaps between them, and the statistics of the labelled gaps obey a closed linear master equation. It gives the path error and the output statistics of the parallel sampler exactly, for any number of steps $N$, sequence length $L$ and schedule. For large $N$, the path error reduces to the thermodynamic metric found for general data in~\cite{LavenantZanella2025}, which yields optimal schedules as geodesics. Since the path error is known exactly at any $N$, we can test the large-$N$ theory directly and identify when it applies: the uniform schedule requires $N\gg\xi$, where $\xi$ is the correlation length, while the cosine schedule of~\cite{Chang2022} only requires $N\gg\sqrt\xi$. We compute the quality of the generated text, measured by its generative perplexity, exactly. For data with long correlations it deteriorates even when the path error saturates, and the geodesic schedule improves it substantially, although it was constructed only to minimize the path error. For the short-range correlations of English text, by contrast, the optimal schedule is nearly uniform.

\section{Parallel sampling error as entropy production}\label{sec:ep}

MDMs generate text by reversing a forward masking process that sequentially masks a sequence of $L$ words (tokens). Training a neural network to learn the masking process on samples of realistic text subsequently allows the generation of new text by reversing the process starting from a sequence of $L$ fully masked tokens. While for realistic text the data distribution is not known and the reverse process is only approximate due to training, the setting considered here removes both restrictions. Sequences $\vec{x}=(x_1,\dots,x_L)$ of tokens from a vocabulary of $V$ symbols are drawn from a distribution $p_{\rm data}$. The forward (masking) process is defined by a schedule $1=p_N>p_{N-1}>\dots>p_0=0$ of revealed fractions. In the step from $p_{n+1}$ to $p_n$, each site that is still revealed is masked independently with probability $\Delta p_n/p_{n+1}$, where $\Delta p_n=p_{n+1}-p_n$. The step $\sigma_i\in\{0,1,...,N-1\}$ in which site $i$ is masked is then independent across sites, with $\P(\sigma_i=n)=\Delta p_n$. Generation traverses the schedule in reverse: in step $n$ the revealed fraction rises from $p_n$ to $p_{n+1}$, and each site that is still masked is revealed with probability $q_n=\Delta p_n/(1-p_n)$, so that the sites revealed in step $n$ are again those with $\sigma_i=n$. Once revealed, a token never changes. A complete history of tokens (``path") is then specified by $(\vec{x},\bsig)$ containing the token $x_i$ and the associated step $\sigma_i$ at which it is masked/revealed. The associated path probability is $p_{\rm data}(\vec{x})\,w(\bsig)$ with $w(\bsig)=\prod_i\Delta p_{\sigma_i}$.

We assume knowledge of the exact conditional distribution $p_{\rm data}(x_i| \vec{x}_S)$ of each masked site $i$ conditional on the set $S$ of revealed sites~\cite{Ou2025}, where $\vec{x}_S=\{x_j:j\in S\}$. This distribution is also called the {\it ideal denoiser} and corresponds to perfect training, since it isolates the error of parallel decoding from the error of learning. Two generative processes are compared:
\begin{itemize}
\item The {\it exact sampler} is precisely the time reversal of masking, characterized by the path probability $P(\vec{x},\bsig)=p_{\rm data}(\vec{x})\,w(\bsig)$. At step $n$, this would require unmasking sites with the conditional probability $p_{\rm data}(\vec{x}_{U_n}| \vec{x}_{S_n})$, where $U_n$ denotes the sites revealed at step $n$ and $S_n$ the sites revealed before step $n$. Thus we can also write
\be
\label{Pprob}
P(\vec{x},\bsig)=w(\bsig)\prod_{n=0}^{N-1}p_{\rm data}(\vec{x}_{U_n}| \vec{x}_{S_n})
\ee

\item The parallel sampler approximates the exact conditional of the jointly revealed sites by a product
\be
p_{\rm data}(\vec{x}_{U_n}| \vec{x}_{S_n})\approx \prod_{i\in U_n}p_{\rm data}(x_i| \vec{x}_{S_n})
\ee
Thus the path probability is
\be
\label{Qprob}
Q(\vec{x},\bsig)=w(\bsig)\prod_{n=0}^{N-1}\prod_{i\in U_n}p_{\rm data}(x_i| \vec{x}_{S_n})
\ee
\end{itemize}
The exact sampler reveals the same groups as the parallel sampler, but sampling it with a denoiser requires drawing each group token by token via the chain rule, at the cost of one network evaluation per token. Thus the two samplers coincide if tokens are revealed one at a time.

The relative error of the parallel sampler compared with the exact sampler is characterised by the relative entropy (Kullback--Leibler divergence)
\begin{align}
B&=\sum_{\vec{x},\bsig}P(\vec{x},\bsig)\ln\frac{P(\vec{x},\bsig)}{Q(\vec{x},\bsig)}=\left<\ln \frac{P(\vec{X},\bsig)}{Q(\vec{X},\bsig)}\right>
\label{eq:B}
\end{align}
We will generally use $\langle...\rangle$ for expectations with respect to the exact process, and $\langle...\rangle_Q$ for the parallel sampled process. 
Since $P(\vec{x},\bsig)$ is likewise the path probability of the forward process, Eq.~\eqref{eq:B} is just the \emph{entropy production} of stochastic thermodynamics~\cite{KPB2007,Seifert2012}. $B$ vanishes when generation exactly reverses masking, and it is positive otherwise. With Eqs.~\eqref{Pprob},\eqref{Qprob} Eq.~\eqref{eq:B} becomes a sum over steps
\begin{equation}
B=\sum_{n=0}^{N-1}\left\langle\ln\frac{p_{\rm data}(\vec{X}_{U_n}| \vec{X}_{S_n})}
{\prod_{i\in U_n}p_{\rm data}(X_i| \vec{X}_{S_n})}\right\rangle
\label{eq:B2}
\end{equation}
Each term compares, for the tokens revealed in step $n$, their exact joint probability given everything revealed before with the product of their individual probabilities. Replacing a joint distribution by the product of its single-site marginals is a mean-field approximation, so $B$ is the accumulated error of one mean-field approximation per step. By Gibbs' inequality each term is non-negative, and it vanishes when the tokens revealed together are independent given the revealed ones, in particular when at most one token is revealed. In Appendix~\ref{sec:mdmloss}, the relationship between $B$ and the variational bound that MDMs are trained to minimize~\cite{Sahoo2024,Shi2024} is discussed: the variational bound measures the quality of the denoiser, whereas $B$ measures the additional error incurred when generation uses finitely many steps.

\paragraph*{Factorising a single step.}
Each term of Eq.~\eqref{eq:B2} can be taken apart further. Label the sites revealed in step $n$ as $i_1,\dots,i_m$, in any order. The chain rule then factorises their exact joint distribution as if they were drawn one after the other, each token conditioned on the tokens drawn before it
\begin{equation}
p_{\rm data}(\vec{x}_{U_n}| \vec{x}_{S_n})=\prod_{j=1}^mp_{\rm data}(x_{i_j}| x_{i_1},\dots,x_{i_{j-1}};\vec{x}_{S_n})
\label{eq:chain}
\end{equation}
The parallel sampler uses $\prod_{j=1}^mp_{\rm data}(x_{i_j}| \vec{x}_{S_n})$ instead, i.e., every token is drawn as if it were the first one. The ratio of the two is a product of single-token ratios, so the logarithm in Eq.~\eqref{eq:B2} becomes a sum
\begin{align}
&\left\langle\ln\frac{p_{\rm data}(\vec{X}_{U_n}| \vec{X}_{S_n})}{\prod_{j=1}^mp_{\rm data}(X_{i_j}| \vec{X}_{S_n})}\right\rangle\nonumber\\
&=\sum_{j=2}^m\left\langle\ln\frac{p_{\rm data}(X_{i_j}| X_{i_1},\dots,X_{i_{j-1}};\vec{X}_{S_n})}{p_{\rm data}(X_{i_j}| \vec{X}_{S_n})}\right\rangle\nonumber\\
&=\sum_{j=2}^mI_{S_n}\big(i_j;\{i_1,\dots,i_{j-1}\}\big)
\label{eq:steporder}
\end{align}
The $j$-th term is the average gain in log-probability of token $i_j$ from knowing the co-revealed tokens drawn before it, which the parallel sampler discards. The first token gains nothing, since nothing co-revealed precedes it. In the last line we have introduced the \emph{mutual information} between token $i_j$ and tokens $\{i_1,\dots,i_{j-1}\}$. For two single sites we write $I_S(i;j)$, which is given by
\be
I_S(i;j)=\left\langle\ln\frac{p_{\rm data}(X_i,X_j| \vec{X}_{S})}{p_{\rm data}(X_{i}| \vec{X}_{S})p_{\rm data}(X_{j}| \vec{X}_{S})}\right\rangle
\label{mipair}
\ee
which follows from the $j=2$ term of Eq.~\eqref{eq:steporder} with $p(x_i,x_j|\vec{x}_S)=p(x_i|x_j,\vec{x}_S)p(x_j|\vec{x}_S)$. This mutual information of pairs is the central quantity in the theory that we develop in the following. Note that the individual terms in Eq.~\eqref{eq:steporder} depend on the order chosen, but their sum does not, because the left-hand side does not refer to any order. The order is therefore a bookkeeping device that can be chosen for convenience. This freedom is useful whenever the data have a short-range structure, such as for Markov data, which we discuss next.

\section{Markov data and shielding}\label{sec:markov}

\subsection{Markov data}
Equation~\eqref{eq:B2} holds for any data, but evaluating it requires the joint statistics of all tokens revealed together, given everything revealed before. This becomes tractable for data generated by a stationary Markov chain, in which each token depends only on its predecessor. The chain is specified by a $V\times V$ transition matrix $T$, where $V$ is the number of token types and
\begin{equation}
T_{bd}=\P(x_{i+1}=d| x_i=b),\qquad \sum_dT_{bd}=1,
\end{equation}
and by its stationary distribution $\pi$, the normalised solution of $\sum_b\pi_bT_{bd}=\pi_d$. A sequence then has probability
\begin{equation}
\label{pdatamarkov}
p_{\rm data}(\vec{x})=\pi_{x_1}\,T_{x_1x_2}\,T_{x_2x_3}\cdots T_{x_{L-1}x_L}.
\end{equation}
The probability of going from token $b$ to token $d$ in $k$ steps is the matrix element $(T^k)_{bd}$ of the $k$-th matrix power, and $(T^k)_{bd}\to\pi_d$ for large $k$, when the memory of the starting token is lost. The transition matrix also satisfies
\be
\label{matrixnorm}
\sum_v (T^k)_{av}\,(T^{r-k})_{vc}=(T^r)_{ac},\qquad r\ge k
\ee
In language modelling this model is the bigram model~\cite{Shannon1948}: for characters of English text $V=27$, for words $V\sim10^4$--$10^5$.

The simplest example has $V=2$ token types, written as spins $s=\pm1$, which switch with probability $f$ at each step:
\begin{align}
T&=\begin{pmatrix}1-f&f\\ f&1-f\end{pmatrix},\qquad \pi=\big(\tfrac12,\tfrac12\big),\nonumber\\
(T^k)_{uv}&=\tfrac12\big(1+\lambda^ks_us_v\big),
\label{eq:Ising}
\end{align}
with $\lambda=1-2f$. This is the zero-field Ising chain, $p_{\rm data}\propto\exp(J\sum_is_is_{i+1})$ with $\tanh J=\lambda$. Its correlations decay as $\langle s_is_j\rangle=\lambda^{|i-j|}=e^{-|i-j|/\xi}$ with correlation length $\xi=-1/\ln\lambda$, and switches are domain walls, which occur independently with density $f\simeq1/(2\xi)$. We use this chain as the main example in Sec.~\ref{sec:results}.

\paragraph*{Correlation length.}
For a general chain, the approach of $(T^k)_{bd}$ to $\pi_d$ is controlled by the
eigenvalues $1=\mu_0>|\mu_1|\ge|\mu_2|\ge\dots$ of $T$. For a diagonalisable $T$,
$(T^k)_{bd}-\pi_d$ is a sum of terms proportional to $\mu_j^k=e^{-k/\xi_j}$
($j\ge1$), with correlation lengths $\xi_j=-1/\ln|\mu_j|$. The largest of them,
$\xi\equiv\xi_1$, sets the range of the correlations: a token at distance $k\gg\xi$
from a known token is distributed as if that token were unknown. For the spin chain
of Eq.~\eqref{eq:Ising} the only non-trivial eigenvalue is $\lambda$, so
$\xi=-1/\ln\lambda$ as above. Large $\xi$ means predictable data: switches are rare,
with density $f\simeq1/(2\xi)$, and the chain consists of domains of mean length
$1/f\simeq2\xi$. As $f\to\frac12$, $\xi\to0$ and neighbouring spins become
independent.

We will see below that both the range of the correlations (the correlation length) and their strength (measured by the mutual information) matter: the strength sets the size of the error, and the range sets how the error is distributed over the course of generation.

\subsection{Shielding}
The key property of Markov data is that revealed tokens \emph{shield}. We call a maximal sequence of $\ell$ masked sites a gap, and label it $(a,\ell,c)$ by the revealed tokens $a$ and $c$ on its left and right ($a,c=\nul$ at the ends of the sequence). Sites inside a gap are numbered $k=1,\dots,\ell$ from the left, so site $k$ is a distance $k$ from the left boundary and $\ell+1-k$ from the right one. Because the probability of a sequence is a product of nearest-neighbour factors, fixing the revealed tokens cuts it into independent pieces: the contents of different gaps are independent, and the content of a gap depends only on its two boundary tokens. In particular, the probability $m^{(a,\ell,c)}_k(v)$ that site $k$ of the gap carries token $v$ is the probability of going from $a$ to $v$ in $k$ steps and from $v$ to $c$ in the remaining $\ell+1-k$ steps, normalised by the probability of going from $a$ to $c$ in $\ell+1$ steps:
\begin{align}
m^{(a,\ell,c)}_k(v)&=\frac{(T^k)_{av}\,(T^{\ell+1-k})_{vc}}{(T^{\ell+1})_{ac}}
\label{eq:marg}
\end{align}
where the normalisation is fixed by Eq.~\eqref{matrixnorm}
\be
\sum_vm^{(a,\ell,c)}_k(v)=1
\ee
At the ends of the sequence the missing boundary simply drops out, which we implement by setting $(T^m)_{\nul v}\equiv\pi_v$, $(T^m)_{v\nul}\equiv1$ and $(T^m)_{\nul\nul}\equiv1$. The ideal denoiser $p_{\rm data}(x_i| \vec{x}_S)$ is then precisely $m^{(a,\ell,c)}_k(v)$: the probability that site $k$ assumes token $v$ inside the gap $(a,\ell,c)$, where the conditional on $S_n$ is reduced to the dependence on $(a,\ell,c)$ due to shielding. Because $(T^k)_{av}\to\pi_v$ for $k\gg\xi$, a boundary token influences only the
sites of a gap within a few correlation lengths of it; deeper inside a long gap,
$m^{(a,\ell,c)}_k(v)\to\pi_v$.

We can also express $p_{\rm data}(x_i,x_j| \vec{x}_S)$ which the exact process uses to reveal two co-revealed sites. Denoting $m^{(a,\ell,c)}_{k,k'}(v,u)$ the joint probability that that site $k$ assumes token $v$ and site $k'$ assumes token $u$ inside gap $(a,\ell,c)$ we have
\be
m^{(a,\ell,c)}_{k,k'}(v,u)=\frac{(T^{k})_{av}\,(T^{k'-k})_{vu}\,(T^{\ell+1-k'})_{uc}}{(T^{\ell+1})_{ac}}
\label{eq:marg2}
\ee
with $\sum_um^{(a,\ell,c)}_{k,k'}(v,u)=m^{(a,\ell,c)}_k(v)$. 

\paragraph*{Path error for Markov data.}
For Markov data, shielding simplifies Eq.~\eqref{eq:steporder} further. Order the co-revealed sites from left to right, so that $i_{j-1}$ is the nearest co-revealed site to the left of $i_j$. By the chain rule for the mutual information,
\be
\label{mishield}
I_{S_n}(i_j;\{i_1,\dots,i_{j-1}\})&=&I_{S_n}\big(i_j;\{i_1,\dots,i_{j-2}\}| i_{j-1}\big)\nonumber\\
&&+I_{S_n}(i_j;i_{j-1})
\ee
and the first term vanishes, since given $X_{i_{j-1}}$ and $\vec{X}_{S_n}$ the token $X_{i_j}$ is independent of everything further to the left. Every term in Eq.~\eqref{eq:steporder} thus reduces to a pair mutual information, $I_{S_n}(i_j;i_{j-1})$; it vanishes if a site of $S_n$ lies between the two, i.e., if they belong to different gaps. Three co-revealed tokens at sites $k<k'<k''$ in the same gap therefore cost $I_{S_n}(k;k')+I_{S_n}(k';k'')$, irrespective of their spacing. The outer pair is correlated, $I_{S_n}(k;k'')>0$, but its correlation is captured by the middle token and is already counted. Drawing from right to left gives the same two pairs, since the mutual information is symmetric. Using Eqs.~\eqref{eq:steporder},\eqref{mishield}, Eq.~\eqref{eq:B2} therefore becomes
\be
B=\sum_{n=0}^{N-1}\Big\langle\sum_{\substack{i<i'\in U_n\\ \text{consecutive}}}I_{S_n}(i;i')\Big\rangle
\label{eq:B3}
\ee
where the inner sum runs over pairs of consecutive co-revealed sites with no co-revealed site between them. If sites $i<i'$ lie in a gap $(a,\ell,c)$ of $S_n$, at positions $k<k'$ within it, shielding implies that their mutual information given $X_{S_n}$ depends only on the gap. We write $I_{S_n}(i;i')=I^{(a,\ell,c)}_{kk'}$ and Eq.~\eqref{mipair} leads to the expression
\be
I^{(a,\ell,c)}_{kk'}\equiv\sum_{v,u}m^{(a,\ell,c)}_{k,k'}(v,u)\,\ln\frac{m^{(a,\ell,c)}_{k,k'}(v,u)}{m^{(a,\ell,c)}_k(v)\,m^{(a,\ell,c)}_{k'}(u)}
\label{eq:Igap}
\ee
while $I_{S_n}(i;i')=0$ if they lie in different gaps. It remains to average Eq.~\eqref{eq:B3} over the sets $U_n$ and $S_n$ for which we need the gap statistics.

\section{The analogy with RSA}\label{sec:rsa}
\begin{figure}[t]
\centering
\resizebox{\columnwidth}{!}{%
\begin{tikzpicture}[x=1cm,y=1cm,>=Stealth]
\def\w{0.62}
\node[anchor=west] at (-0.6,1.05) {(a)};
\fill[cplus!80] (0,0) rectangle +(0.56,0.56); \node[white] at (0.28,0.28) {$a$};
\foreach \i in {1,...,8}{\fill[cmask] (\i*\w,0) rectangle +(0.56,0.56); \node[gray] at (\i*\w+0.28,0.28) {\small$\M$};}
\fill[cminus!85] (9*\w,0) rectangle +(0.56,0.56); \node[white] at (9*\w+0.28,0.28) {$c$};
\draw[very thick] (3*\w,0) rectangle +(0.56,0.56); \draw[very thick] (6*\w,0) rectangle +(0.56,0.56);
\draw[->] (3*\w+0.28,1.0)--(3*\w+0.28,0.62); \draw[->] (6*\w+0.28,1.0)--(6*\w+0.28,0.62);
\node at (3*\w+0.28,-0.27) {\small$k$}; \node at (6*\w+0.28,-0.27) {\small$k'$};
\draw[decorate,decoration={brace,amplitude=3pt,mirror}] (1*\w,-0.48)--(2*\w+0.56,-0.48) node[midway,below=2pt] {\scriptsize$(a,k\!-\!1,v)$};
\draw[decorate,decoration={brace,amplitude=3pt,mirror}] (4*\w,-0.48)--(5*\w+0.56,-0.48) node[midway,below=2pt] {\scriptsize$(v,k'\!-\!k\!-\!1,u)$};
\draw[decorate,decoration={brace,amplitude=3pt,mirror}] (7*\w,-0.48)--(8*\w+0.56,-0.48) node[midway,below=2pt] {\scriptsize$(u,\ell\!-\!k',c)$};
\begin{scope}[yshift=-2.25cm]
\node[anchor=west] at (-0.6,1.05) {(b)};
\newcommand{\cell}[3]{%
  \ifx#3p\fill[cplus] (#1*\w,#2) rectangle +(0.56,0.56); \node[white] at (#1*\w+0.28,#2+0.28) {\small$+$};\fi
  \ifx#3m\fill[cminus] (#1*\w,#2) rectangle +(0.56,0.56); \node[white] at (#1*\w+0.28,#2+0.28) {\small$-$};\fi
  \ifx#3q\fill[cmask] (#1*\w,#2) rectangle +(0.56,0.56); \node[gray] at (#1*\w+0.28,#2+0.28) {\small$\M$};\fi}
\cell{0}{0.25}{p}\cell{1}{0.25}{q}\cell{2}{0.25}{p}\cell{3}{0.25}{q}\cell{4}{0.25}{m}\cell{5}{0.25}{q}\cell{6}{0.25}{m}
\node[anchor=west,font=\small] at (7*\w,0.53) {joint: 1 switch};
\cell{0}{-0.55}{p}\cell{1}{-0.55}{q}\cell{2}{-0.55}{m}\cell{3}{-0.55}{q}\cell{4}{-0.55}{p}\cell{5}{-0.55}{q}\cell{6}{-0.55}{m}
\node[anchor=west,font=\small] at (7*\w,-0.27) {independent: 3 switches};
\end{scope}
\end{tikzpicture}}
\caption{(a) A labelled gap $(a,\ell,c)$ of masked sites $\M$. Revealing the selected sites $k<k'$ with tokens $v,u$ splits it into three conditionally independent labelled gaps, as a deposition does in RSA. (b) The late-stage error mechanism for data with a single correlation length (binary tokens $\pm$). A gap whose boundaries disagree contains one switch (domain wall). Two sites revealed in the same step can be drawn in the impossible order $(-,+)$, which creates two spurious switches that would not occur under exact sampling.}
\label{fig:schematic}
\end{figure}
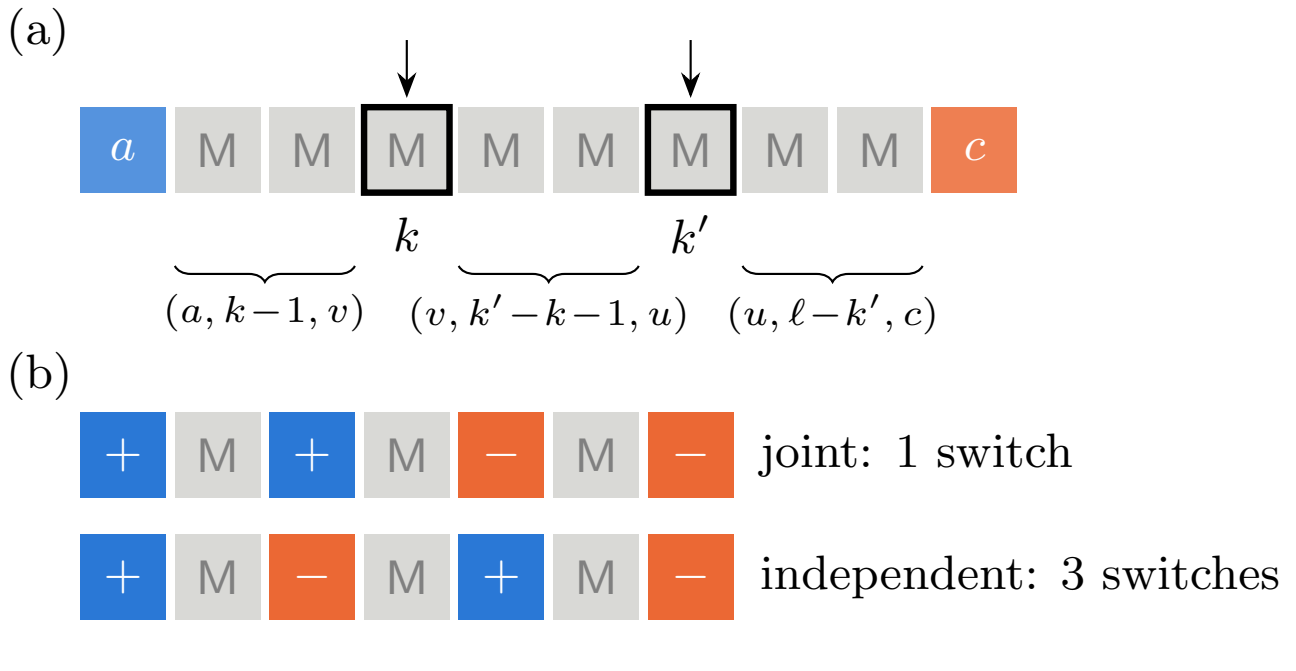

Shielding is also what makes one-dimensional RSA exactly solvable~\cite{Renyi1958,Evans1993}. In R\'enyi's ``car parking'' problem, rods are deposited one by one at random positions on a line, and each deposited rod splits a gap into two gaps that evolve independently afterwards. The density of gaps of each length therefore obeys a closed linear equation, even though the full configuration is a complicated many-body state. In masked generation, revealed tokens are the deposited particles and masked sequences are the gaps, see Fig.~\ref{fig:schematic}(a). The value of a revealed token is a \emph{species} that labels the gaps on either side, just as particle orientations label gaps in the ``Paris car parking'' problem of anisotropic particles~\cite{Baule2017}. The species of a newly deposited particle is drawn from Eq.~\eqref{eq:marg}. Apart from the factor $\pi_v$, its numerator is a product of two factors, $(T^k)_{av}/\pi_v$ and $(T^{\ell+1-k})_{vc}/\pi_c$, each describing the influence of one neighbouring revealed token, and each decays to $1$ (no influence) beyond the correlation lengths of the chain. This is the factorised form $\Omega(x,y)\propto\omega(x)\,\omega(y-x)$ of the deposition rate in interacting RSA, which was solved exactly in~\cite{Baule2019}. Here, however, the interaction selects the species of the deposited particle rather than whether it is accepted. The new ingredient is the \emph{parallel} update: particles deposited in the same step do not see each other, so their species are chosen as if the others were absent.

\paragraph*{Master equation.}
Let $\mathcal N_n(a,\ell,c)$ be the expected number of gaps of type $(a,\ell,c)$ in the sampler after $n$ steps; initially there is a single gap, $\mathcal N_0(\nul,L,\nul)=1$. In step $n$, every masked site is selected independently with probability $q_n$, and each selected site draws its token from $m^{(a,\ell,c)}_k(v)$. A parent gap $(a',\ell',c')$ therefore produces a child gap $(a,\ell,c)$ in one of three ways:
\begin{itemize}
\item \emph{Survival:} none of its sites is selected. This happens with probability $(1-q_n)^{\ell'}$ and the child is the parent itself.
\item \emph{End child:} one site is selected. The child occupies sites $1,\dots,\ell$ of a parent with the same left boundary $a'=a$. Its $\ell$ sites must remain unselected, with probability $(1-q_n)^\ell$, and site $\ell+1$ must be selected, with probability $q_n$, and receive the token $c$, with probability $m^{(a,\ell',c')}_{\ell+1}(c)$. The mirror-image process creates a child at the right end of a parent with $c'=c$.
\item \emph{Interior child:} two sites are selected. The child occupies sites $k+1,\dots,k+\ell$ of the parent. Its two bounding sites $k$ and $k+\ell+1$ must both be selected, with probability $q_n^2$, its own interior sites must remain unselected, with probability $(1-q_n)^\ell$, and the bounding sites receive tokens $a$ and $c$. The latter are selected with probability $m^{(a',\ell',c')}_{k,k+\ell+1}(a,c)$ for the exact sampler, while the parallel sampler approximates
\be
\label{mprod}
m^{(a',\ell',c')}_{k,k+\ell+1}(a,c)\approx m^{(a',\ell',c')}_k(a)\,m^{(a',\ell',c')}_{k+\ell+1}(c)
\ee
\end{itemize}
Since gaps evolve independently, with statistics that depend only on their labels, the expected number of children follows by summing these probabilities over all parents. This gives the closed linear master equation
\begin{align}
&\mathcal N_{n+1}(a,\ell,c)=(1-q_n)^\ell\Big[\mathcal N_n(a,\ell,c)\nonumber\\
&\qquad+q_n\sum_{\ell'>\ell}\sum_{c'}\mathcal N_n(a,\ell',c')\,m^{(a,\ell',c')}_{\ell+1}(c)\nonumber\\
&\qquad+q_n\sum_{\ell'>\ell}\sum_{a'}\mathcal N_n(a',\ell',c)\,m^{(a',\ell',c)}_{\ell'-\ell}(a)\nonumber\\
&\qquad+q_n^2\sum_{\ell'\ge\ell+2}\sum_{a',c'}\mathcal N_n(a',\ell',c')\nonumber\\
&\qquad\qquad\times\sum_{k=1}^{\ell'-\ell-1}m^{(a',\ell',c')}_{k,k+\ell+1}(a,c)\Big],
\label{eq:master}
\end{align}
where the parallel sampler uses Eq.~\eqref{mprod} for the last term and $m(\nul)\equiv0$, since a revealed token is never $\nul$. The four terms are survival, the left and right end children, and the interior children. Eq.~\eqref{eq:master} is exact for any step size, including a single step in which all sites are revealed. Any number of sites of a parent can be selected in one step, but every child is bounded by at most two of them, so higher powers of $q_n$ cannot occur: a gap has only two ends. The equation is exact for the expected counts even though the configuration generated by the parallel sampler is \emph{not} distributed like the data, because only linearity of expectation is used.

Remarkably, the exact sampler described by Eq.~\eqref{eq:master} has a simple solution. Consider a gap $(a,\ell,c)$. In the data, two tokens $a$ and $c$ at distance $\ell+1$ occur with probability $\pi_a(T^{\ell+1})_{ac}$. Before step $n$ both sites have been revealed with probability $p_n^2$, and the $\ell$ sites between them are still masked with probability $(1-p_n)^\ell$. Since a gap of length $\ell$ can be placed in $L-\ell-1$ positions, this suggests the ansatz
\be
\label{Nexact}
\mathcal N_n(a,\ell,c)=(L-\ell-1)p_n^2(1-p_n)^\ell\pi_a(T^{\ell+1})_{ac}
\ee 
The two edge gaps follow in the same way, $\mathcal N_n(\nul,\ell,c)=p_n(1-p_n)^\ell\pi_c$ and $\mathcal N_n(a,\ell,\nul)=p_n(1-p_n)^\ell\pi_a$, and the fully masked sequence has probability $(1-p_n)^L$. Eq.~\eqref{Nexact} is indeed the solution of Eq.~\eqref{eq:master} for any $n$, as can be shown by direct substitution. This simple result highlights that the exact sampler preserves the data distribution at any step. The parallel sampler on the other hand deviates from it and has no simple solution. The resulting error is measured by $B$, Eq.~\eqref{eq:B2}. In the physics context, the exact sampler preserves the equilibrium distribution, while the parallel sampler is a non-equilibrium process giving rise to the dissipation $B$.

\section{Path error and output quality}

\subsection{Path error}

For Markov data we can now evaluate exactly: (i) the path error $B$ of Eq.~(\ref{eq:B2}); and (ii) the {\it generative perplexity} of the output, which measures the output quality of MDMs~\cite{Ou2025}. The average over $U_n,S_n$ in Eq.~\eqref{eq:B3} is taken over the exact process and can be calculated as follows. Given $S_n$, each masked site belongs to $U_n$ independently with probability $q_n$, so two sites $k<k'$ of a gap form a consecutive co-revealed pair with probability $q_n^2(1-q_n)^{k'-k-1}$: both are selected and none of the $k'-k-1$ sites between them is. Averaging over $S_n$ then replaces the sum over the gaps of $S_n$ by the gap densities. Hence
\begin{align}
B=&\sum_{n=0}^{N-1}\sum_{(a,\ell,c)}\mathcal N_{n}(a,\ell,c)\nonumber\\
&\times\sum_{1\le k<k'\le\ell}q_n^2(1-q_n)^{k'-k-1}\,I^{(a,\ell,c)}_{kk'}
\label{eq:Bexact}
\end{align}
with $I^{(a,\ell,c)}_{kk'}$ given by Eq.~\eqref{eq:Igap}.

\subsection{Generative perplexity}

This quantity measures the quality of the final output of the sampler, i.e., after every gap has been filled. After the last step every gap has length zero, and $\mathcal P_{bd}=\mathcal N_N(b,0,d)$ is the expected number of adjacent pairs $(b,d)$ in the output; in RSA terms, the final distribution of nearest-neighbour species. This is enough to quantify how ``typical'' the generated sequences are under the true distribution. In machine learning the average
\be
\CE_Q=-\left<\ln p_{\rm data}(\vec{X})\right>_Q
\ee
is called the \emph{cross-entropy} of the text $\vec{X}$ under the parallel sampler $Q$. Its exponential per token, $e^{\CE_Q/L}$, is the \emph{generative perplexity}, an effective number of equally likely choices per token, which is the standard quality measure for MDMs~\cite{Lou2024,Sahoo2024,Ou2025}. In practice it is evaluated with a large reference model in place of the unknown $p_{\rm data}$, and it can be lowered artificially by reducing the diversity of the output~\cite{Zheng2025}; for Markov data neither issue arises, since $p_{\rm data}$ is known. For Markov data $\CE_Q$ can be evaluated exactly in the RSA approach. We have from Eq.~\eqref{pdatamarkov}
\be
\ln p_{\rm data}(\vec{x})=\ln\pi_{x_1}+\sum_{i=1}^{L-1}\ln T_{x_ix_{i+1}}
\ee
and thus
\be
\left<\ln p_{\rm data}(\vec{X})\right>_Q=\sum_d \phi_d\ln\pi_d+\sum_{b,d}\mathcal P_{bd}\ln T_{bd}
\ee
Averaging over the output of the sampler, each pair $(b,d)$ contributes $\ln T_{bd}$ times its expected number $\mathcal P_{bd}$, and the first token contributes $\ln\pi_d$ with probability $\phi_d=\mathcal N_N(\nul,0,d)$. These counts depend on the number of steps and on the schedule, which are provided by the master equation. For the exact sampler, $\mathcal P_{bd}=(L-1)\pi_bT_{bd}$, since there are $L-1$ bonds and each bond has $(b,d)$ configuration with probability $\pi_bT_{bd}$. We also have $\phi_d=\pi_d$. For the parallel sampler, $\mathcal P_{bd}$ and $\phi_d$ are given from the gap statistics of the master equation. The difference in cross entropy between exact and parallel sampler is therefore
\begin{align}
\Delta\CE&\equiv\CE_Q+\left<\ln p_{\rm data}(\vec{X})\right>\nonumber\\
&=\sum_{b,d}\big[(L-1)\pi_bT_{bd}-\mathcal{P}_{bd}\big]\ln T_{bd}\nonumber\\
&\quad+\sum_d(\pi_d-\phi_d)\ln\pi_d.
\label{eq:CE}
\end{align}
It counts the excess of improbable adjacent pairs in the output, each weighted by $-\ln T_{bd}$. For the spin chain this is literally the excess energy of spurious domain walls (Sec.~\ref{sec:results}). $\Delta\CE$ thus requires the statistics of $\mathcal{P}_{bd},\phi_d$ obtained from the master equation of the parallel process, while the path error Eq.~\eqref{eq:Bexact} requires only the gap density of the exact sampler, Eq.~\eqref{Nexact}. 

\subsection{Numerical validation}
The master equation~\eqref{eq:master} and Eq.~\eqref{eq:Bexact} are exact, and we
checked them against two computations that do not use the gap representation (see Appendix~\ref{app:numerics} for more details):
(i)~\emph{Exact enumeration.} For short sequences all generation histories can be
listed: each site has one of $N$ reveal steps, and the sequence one of $V^L$
values. For every history, $Q(\vec x,\bsig)$ follows from Eq.~\eqref{Qprob}, with
the ideal denoiser obtained by conditioning $p_{\rm data}$ directly. Summing over
histories with a prescribed schedule $p_n$ gives the expected pair counts and the path relative entropy~\eqref{eq:B}.
For binary chains with $\xi=2,5,20$ ($L=8$, $N\le4$) and a random three-state chain
($L=6$, $N\le3$), the master equation agrees with enumeration to $2\times10^{-14}$, and Eq.~\eqref{eq:Bexact} with the
enumerated $B$ to a relative $2\times10^{-15}$, i.e., to machine precision.
(ii)~\emph{Direct simulation.} For $L=1024$ we simulate the parallel sampler:
in each step the selected sites draw their tokens from conditionals computed from
the state before the step. With $10^4$ samples per case, the simulated $\Delta\CE$
and all pair counts agree with the master equation within two standard errors, and
a simulation of the exact sampler gives $\Delta\CE$ consistent with zero.

\section{Optimal schedules as geodesics}\label{sec:geo}

We now make an additional approximation in Eq.~\eqref{eq:Bexact} by setting $(1-q_n)^{k'-k-1}\approx 1$. This neglects \emph{screening}: a pair at distance $d=k'-k$ is counted in Eq.~\eqref{eq:Bexact} only if none of the $d-1$ sites between them is revealed in
the same step, which fails with probability $\simeq q_n(d-1)$. The pairs that carry
information have $d\lesssim\xi$, since beyond a correlation length two sites are
nearly independent, and $d\lesssim1/p_n$, since they must lie in the same gap. The
approximation therefore requires that few correlated partners of a site are revealed
in the same step, $q_n\min(\xi,1/p_n)\ll1$. It is most restrictive in the first
step, where $p_0=0$ and it requires $\Delta p_0\ll1/\xi$. For $\xi\lesssim1$ the
information is carried by adjacent pairs, for which the screening factor is exactly
one. Eq.~\eqref{eq:Bexact} then becomes
\be
B=\sum_{n=0}^{N-1}q_n^2\sum_{(a,\ell,c)}\mathcal N_{n}(a,\ell,c)\sum_{1\le k<k'\le\ell}I^{(a,\ell,c)}_{kk'}
\label{eq:Bn}
\ee
Substituting Eq.~\eqref{Nexact}, we can use $L-\ell-1\approx L$ (terms with large $\ell$ are suppressed by the factor $(1-p_n)^\ell$) and neglect the edge gaps and the fully masked sequence, whose contributions are of relative order $1/(Lp_n)$. With $q_n=\Delta p_n/(1-p_n)$ we then obtain
\be
\label{Bgeo}
B&=&L\sum_{n=0}^{N-1}\Delta p_n^2\,G(p_n)\\
G(p_n)&=&\frac{p_n^2}{(1-p_n)^2}\sum_{(a,\ell,c)}(1-p_n)^\ell\pi_a(T^{\ell+1})_{ac}\nonumber\\
&&\times\sum_{1\le k<k'\le\ell}I^{(a,\ell,c)}_{kk'}\label{Gexp}
\ee
Eq.~\eqref{Bgeo} highlights that $B$ is proportional to $\Delta p_n^2$ and a term $G(p_n)$ that depends on the data.

As shown in Appendix~\ref{app:Gexp2}, $G(p_n)$ can be evaluated further by performing the sum over the tokens $(a,c)$. Writing the pair in a gap as $l=k$, $d=k'-k$, $r=\ell+1-k'$, Eq.~\eqref{Gexp} becomes
\be
\label{Gexp2}
G(p_n)=\sum_{l,d,r\ge1}p_n^2(1-p_n)^{l+d+r-3}\,\bar I(l,d,r)
\ee
where
\begin{equation}
\label{Ibar}
\bar I(l,d,r)=I(d)-I(d+l)-I(d+r)+I(d+l+r)
\end{equation}
and $I(m)$ is the mutual information of two data tokens at distance $m$
\be
I(m)=\sum_{u,v}\pi_u(T^m)_{uv}\ln\big[(T^m)_{uv}/\pi_v\big]
\ee

\subsection{Geodesic and minimum error}

For a smooth schedule $p_n=p(n/N)$ the steps are $\Delta p_n=\dot p(\tau_n)/N+O(N^{-2})$, with $\tau_n=n/N$ a reduced time. The error per token $B/L$ with Eq.~\eqref{Bgeo} thus becomes for large $N$
\begin{equation}
\frac BL\simeq\frac1N\int_0^1\dot p(\tau)^2\,G(p(\tau))\,d\tau
\label{eq:Bcont}
\end{equation}
where $p(0)=0$ and $p(1)=1$. Eq.~\eqref{eq:Bcont} has the form of the dissipation of a slowly driven system~\cite{SalamonBerry1983,SivakCrooks2012}. When a control parameter $\theta(t)$ is changed over a time $\tau_{\rm p}$, the excess work is $\simeq\tau_{\rm p}^{-1}\int\dot\theta^2\zeta(\theta)$, with $\zeta$ a friction coefficient. Here $G$ plays the role of the friction, the revealed fraction $p$ that of the control parameter, and the number of steps $N$ that of the protocol duration. Analogous speed--accuracy relations have been derived for continuous diffusion models~\cite{Ikeda2025}. An equation in the form of Eq.~\eqref{eq:Bcont} has been derived for general data in~\cite{LavenantZanella2025}, where $G(p)$ is related to the conditional entropy of a masked token when every other site is revealed independently with probability $p$. Eq.~\eqref{Gexp2} is the explicit result for this metric in the case of Markov data.

Minimising Eq.~\eqref{eq:Bcont} at fixed $N$ is a geodesic problem \cite{LavenantZanella2025}. Defining the Lagrangian
\be
\mathcal{L}(\dot p,p)=\dot p^2\,G(p)
\label{eq:Lagr}
\ee
the optimal schedule $p^*(\tau)$ minimizes $\int_0^1\mathcal{L}\,d\tau$ subject to $p(0)=0$ and $p(1)=1$, and therefore satisfies the Euler--Lagrange equation
\be
\frac{d}{d\tau}\frac{\partial\mathcal{L}}{\partial\dot p}-\frac{\partial\mathcal{L}}{\partial p}=2\ddot p\,G(p)+\dot p^2\,G'(p)=0
\label{eq:EL}
\ee
Since $\mathcal{L}$ does not depend explicitly on $\tau$, Eq.~\eqref{eq:EL} has a first integral (the Beltrami identity),
\be
\frac{d}{d\tau}\Big(\dot p\,\frac{\partial\mathcal{L}}{\partial\dot p}-\mathcal{L}\Big)=\frac{d}{d\tau}\big(\dot p^2G(p)\big)=\dot p\,\big(2\ddot p\,G+\dot p^2G'\big)=0\nonumber
\ee
The integrand of Eq.~\eqref{eq:Bcont}, i.e., the error per unit of reduced time, is thus conserved along the optimum:
\be
\big(\dot p^*\big)^2\,G(p^*)=c
\label{eq:conserved}
\ee
In discrete form every step contributes the same amount, $L\,\Delta p_n^2\,G(p_n)\simeq Lc/N^2$, to Eq.~\eqref{Bgeo}. We assume that $G$ is continuous and strictly positive on $[0,1]$, and that $p^*$ is continuously differentiable. The constant in Eq.~\eqref{eq:conserved} is then positive: $c=0$ would imply $\dot p^*\equiv0$, contradicting $p^*(0)=0$ and $p^*(1)=1$. Hence $\dot p^*$ never vanishes, and being continuous it cannot change sign, so that $\dot p^*>0$ and
\be
\dot p^*=\sqrt{c}\;G(p^*)^{-1/2}
\label{eq:pdotopt}
\ee
i.e., the optimal schedule takes small steps where $G$ is large. To integrate Eq.~\eqref{eq:pdotopt}, we define the cumulative length
\be
F(p)=\int_0^{p}\sqrt{G(p')}\,dp'
\label{eq:Fdef}
\ee
which is continuously differentiable with $F'(p)=\sqrt{G(p)}>0$ and $F(0)=0$. With Eq.~\eqref{eq:pdotopt},
\be
\frac{d}{d\tau}F\big(p^*(\tau)\big)=\sqrt{G(p^*)}\;\dot p^*=\sqrt{c}
\ee
Integrating over $[0,\tau]$ using $p^*(0)=0$ gives
\be
F\big(p^*(\tau)\big)=\int_0^{p^*(\tau)}\sqrt{G(p')}\,dp'=\sqrt{c}\,\tau
\label{eq:tauint}
\ee
The boundary condition $p^*(1)=1$ fixes the constant to $\sqrt{c}=\Lambda$, where
\begin{equation}
\Lambda=\int_0^1\sqrt{G(p)}\,dp
\label{eq:Lambda}
\end{equation}
is the thermodynamic length~\cite{SalamonBerry1983,Crooks2007}. Since $F$ is strictly increasing, it is invertible, and the optimal schedule is $p^*(\tau)=F^{-1}(\Lambda\tau)$, equivalently
\be
\tau(p^*)=\frac{F(p^*)}{\Lambda}=\frac{1}{\Lambda}\int_0^{p^*}\sqrt{G(p')}\,dp'
\label{eq:tauopt}
\ee
Inserting Eq.~\eqref{eq:conserved} with $c=\Lambda^2$ into Eq.~\eqref{eq:Bcont} yields the error of the optimal schedule \cite{LavenantZanella2025}
\be
\label{opterror}
\frac BL\simeq\frac1N\int_0^1\big(\dot p^*\big)^2G(p^*)\,d\tau=\frac{\Lambda^2}{N}
\ee
which is the global minimum. Indeed, for any monotone schedule with $p(0)=0$ and $p(1)=1$, the Cauchy--Schwarz inequality gives $\Lambda^2=\big(\int_0^1\sqrt{G(p)}\,\dot p\,d\tau\big)^2\le\int_0^1\dot p^2G(p)\,d\tau$, with equality if and only if $\dot p^2G$ is constant, i.e., for the schedule~\eqref{eq:tauopt}. 

Note that the geodesic is optimal only to leading order in $1/N$; at very small $N$ other schedules can do marginally better.
The passage from Eq.~\eqref{Bgeo} to Eq.~\eqref{eq:Bcont} also requires the steps to
be small compared with the scale over which $G$ varies, $\Delta p_n\ll|G/G'|$. This
scale is again set by the correlation length. The typical gap length $1/p$ is
compared with $\xi$: early in generation, $p\lesssim1/\xi$, gaps are longer than a
correlation length and masked sites share information with many others; later,
revealed sites lie within a correlation length of every masked site and screen most
of it. For long correlations $G$ therefore changes on the scale $p\sim1/\xi$, and
both conditions fail if the first steps are wider than $1/\xi$. For the uniform
schedule they amount to $N\gg\xi$. Since $B$ is known exactly for Markov data,
Eq.~\eqref{eq:Bexact}, these conditions can be tested directly
(Sec.~\ref{sec:results}).

\section{Examples}\label{sec:results}

\begin{figure}
\centering
\includegraphics[width=\columnwidth]{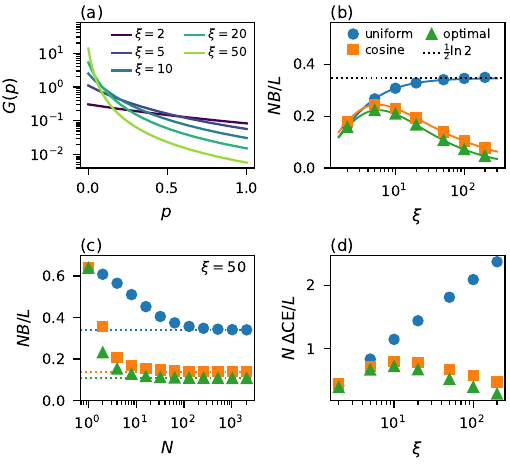}
\caption{Binary chain with $L=4096$ and $N$ up to $2048$. (a) The metric $G(p)$, Eq.~\eqref{Gexp2}, for $\xi=2,5,10,20,50$ (dark to light): for long correlations it is concentrated at small $p$. (b) Path error $NB/L$ against $\xi$ for the uniform, cosine and optimal schedules. Symbols: exact values from Eq.~\eqref{eq:Bexact} at $N=2048$; solid lines: the large-$N$ results $U$, Eq.~\eqref{Uex}, $\int_0^1\dot p^2G\,d\tau$ and $\Lambda^2$ computed from $G$. (c) $NB/L$ against $N$ at $\xi=50$; dotted lines mark the large-$N$ asymptotic results, see (b). (d) Excess cross-entropy per token, $N\Delta\CE/L$, from the master equation~\eqref{eq:master} at $N=2048$, against $\xi$.}
\label{fig:binary}
\end{figure}

\begin{figure}
\centering
\includegraphics[width=\columnwidth]{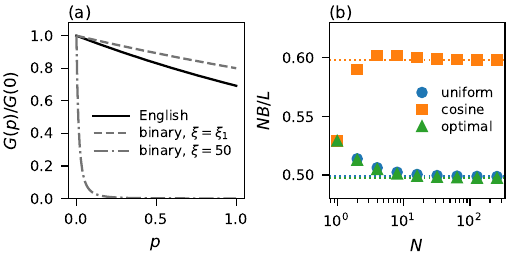}
\caption{Character-level English bigram model for $L=1024$ and $N$ up to $512$. (a) The metric $G(p)$, normalised by $G(0)$, compared with the binary chain at the same correlation length, $\xi=\xi_1=0.76$, and at $\xi=50$. (b) Path error $NB/L$ against $N$ for the uniform, cosine and optimal schedules; dotted lines mark the large-$N$ coefficients $U$, Eq.~\eqref{Uex}, $\int_0^1\dot p^2G\,d\tau$ and $\Lambda^2$ computed from $G$.}
\label{fig:english}
\end{figure}

We evaluate the expressions of the preceding sections for two data sets: the binary chain of Eq.~\eqref{eq:Ising}, whose correlation length $\xi$ can be tuned, and a bigram model of English text, whose correlations are short-ranged. All quantities are computed exactly from the transition matrix $T$; no sampling is involved except in the validation against direct simulations described above. 

Three masking/unmasking schedules are investigated:
\begin{enumerate}
\item[(i)] The \emph{uniform schedule}, where
\be
p(\tau)=\tau
\ee
Eq.~\eqref{eq:Bcont} gives $B/L\simeq U/N$, where the constant $U$ can be calculated in closed form (see Appendix~\ref{app:Uex}):
\be
\label{Uex}
U=\int_0^1G(p)\D p=I(1)-\frac{1}{2}I(2)
\ee
Remarkably, although $G(p)$ involves mutual information $I(m)$ at all lags $m\ge1$, its integral over $p$ retains only the two shortest lags; all contributions from $m\ge3$ cancel exactly. The error of the uniform schedule is therefore determined solely by nearest- and next-nearest-neighbour correlations in the data.
\item[(ii)] The \emph{cosine schedule} of MaskGIT~\cite{Chang2022}, where
\be
\label{cosp}
p(\tau)=1-\cos(\pi\tau/2)
\ee
The large $N$ error follows from Eq.~\eqref{eq:Bcont}.

\item[(iii)] The \emph{optimal schedule}, where $\dot p\propto G^{-1/2}$. We invert Eq.~\eqref{eq:tauopt} for $p(\tau)$ and evaluate at $\tau_n=n/N$. The large $N$ error is given in Eq.~\eqref{opterror}.
\end{enumerate}

Fig.~\ref{fig:binary}(a) shows $G(p)$ for $\xi=2$ to $50$. The metric $G(p)$ is calculated from Eq.~\eqref{Gexp2} on a grid of $p$ with the mutual information $I(m)$ computed from powers of $T$. The sum over gap lengths converges because of the factor $(1-p)^\ell$, and it is truncated once the remaining terms fall below $10^{-15}$. By Eq.~\eqref{Gexp}, $G(p)$ is the information shared by pairs of masked sites in the same gap, given the revealed tokens. For short correlations $G$ varies little with $p$. As $\xi$ grows, $G$ becomes large at small $p$ and decays quickly once the typical distance $1/p$ between revealed sites falls below $\xi$: the metric is concentrated at the start of generation. Early on, gaps are long and a masked spin shares information with all spins within a distance $\sim\xi$. Late in generation, gaps are short compared with $\xi$ and the revealed boundaries already fix most sites inside, so that learning one masked site tells little about another. Only the rare gaps whose boundaries disagree contribute, because they must contain one switch at an uncertain position. Two sites of such a gap that are drawn independently can end up on the wrong sides of the switch, producing spurious switches, see Fig.~\ref{fig:schematic}(b).

Fig.~\ref{fig:binary}(b) shows how the path error depends on the correlation length. Symbols denote the finite-$N$ results, where the exact path error is computed from Eq.~\eqref{eq:Bexact} with the gap densities~\eqref{Nexact}. Solid lines show the large-$N$ result Eq.~\eqref{eq:Bcont}, which agrees perfectly for $N=2048$ used here. For the uniform schedule, $NB/L$ increases with $\xi$ and saturates at $\frac12\ln2$. This saturation follows directly from Eq.~\eqref{Uex}: as $\xi\to\infty$, the binary chain approaches perfect order and both $I(1)$ and $I(2)$ approach $\ln2$, so $U\to\ln2-\frac12\ln2=\frac12\ln2$. The uniform schedule spends most of its steps where $G$ is small, but its first steps cover the region at small $p$ where $G$ is large. The optimal schedule instead takes small steps in that region and large steps afterwards. Its error is largest for $\xi\approx5$ and then decreases: more predictable data make parallel decoding \emph{cheaper}, provided the schedule starts with small steps. The gain over the uniform schedule therefore grows with $\xi$, to a factor of about $7$ at $\xi=200$. The cosine schedule also starts with small steps and follows the optimal schedule closely for $\xi\lesssim20$, but its error falls behind for longer correlations.

Fig.~\ref{fig:binary}(c) shows the approach to the large-$N$ coefficients at $\xi=50$. For $N=1$ all schedules coincide, since all tokens are revealed in one step. The uniform schedule reaches its coefficient only for $N$ of a few hundred, i.e., $N\gg\xi$, as required by the conditions of Sec.~\ref{sec:geo}: its first step, $\Delta p_0=1/N$, must be small compared with $1/\xi$. The cosine and optimal schedules converge already for $N\approx10$--$20$, because their first steps are much smaller. For the cosine schedule, whose first step is $\Delta p_0\simeq\pi^2/(8N^2)$, they amount only to $N\gg\sqrt\xi$. The optimal schedule adapts its first step to the metric: by Eq.~\eqref{eq:tauopt}, $F(p)\simeq\sqrt{G(0)}\,p$ for small $p$, so that $\Delta p_0\simeq\Lambda/(N\sqrt{G(0)})$ and the condition becomes $N\gg\xi\Lambda/\sqrt{G(0)}$. Since $G(0)=\sum_dI(d)$ grows in proportion to $\xi$ while $\Lambda$ decreases, this bound grows only slowly; for the binary chain it increases from $2.2$ at $\xi=5$ to $5.7$ at $\xi=200$.

For the binary chain, Eq.~\eqref{eq:CE} takes a simple form. Since $\pi$ is uniform, the first-token term vanishes, and $T_{bd}$ takes only two values, $1-f$ for equal neighbours and $f$ for a switch, so the expected switch density is exactly $f$. If $r$ is the expected number of switches in the actual output, the bracket in Eq.~\eqref{eq:CE} is $(L-1)f-r$ for switches and $r-(L-1)f$ for equal neighbours, so that per token
\begin{equation}
\frac{\Delta\CE}{L}=\frac{L-1}{L}\,\delta\rho\,\ln\frac{1-f}{f},
\label{eq:CEdefect}
\end{equation}
where $\delta\rho=r/(L-1)-f$ is the excess density of switches in the output. Each spurious switch costs $\ln[(1-f)/f]\simeq\ln(2\xi)$, the energy $2J$ of a domain wall. 

Fig.~\ref{fig:binary}(d) shows the consequence in the perplexity, which is computed from the master equation~\eqref{eq:master} of the parallel sampler, which gives $\mathcal P_{bd}$, $\phi_d$ and hence $\Delta\CE$. For the uniform schedule, $N\delta\rho\to 0.39$ for large $\xi$ as measured from the output, so that by Eq.~\eqref{eq:CEdefect} the perplexity excess grows as $N\Delta\CE/L\simeq0.39\ln(2\xi)$. On the other hand, the error per token $NB/L\to\frac{1}{2}\ln 2$ saturates in this regime (see Fig.~\ref{fig:binary}(b)). The perplexity grows for large $\xi$ because each spurious switch becomes more costly as the data become more predictable. In other words, as correlations get longer, the parallel sampler makes the same number of mistakes per token, but each mistake looks less plausible to the data model because the data become more predictable. For the optimal schedule, which suppresses switches, $N\Delta\CE/L$ instead decreases beyond $\xi\approx10$, and the reduction relative to the uniform schedule grows from a few percent at $\xi=2$ to a factor of about $9$ at $\xi=200$. The schedule that minimizes the path error thus also improves the quality of the output, although it was not constructed to do so.

\subsection{English text example}
To test the theory on real language statistics, we fitted a character-level bigram model of English to a public-domain corpus (Tiny Shakespeare~\cite{Karpathy2015}). The text was lowercased, every character other than a--z was mapped to a space, and runs of spaces were collapsed, leaving $1.06\times10^6$ characters over $V=27$ symbols. The transition matrix was estimated from bigram counts with add-$\alpha$ smoothing, $T_{bd}=(n_{bd}+\alpha)/(n_b+V\alpha)$ with $\alpha=0.01$, and $\pi$ is the stationary distribution of the fitted $T$. Smoothing is needed because the parallel sampler produces pairs that never occur in the corpus; without it $\Delta\CE$ would be infinite.
 
The correlations of this chain are short-ranged, with the largest correlation length only $\xi_1=0.76$ characters, but strong: adjacent characters share $I(1)=0.53$ nats of information, as in ``q'' and ``u''. The metric is correspondingly nearly flat, see Fig.~\ref{fig:english}(a), falling only from $G(0)=0.60$ to $G(1)=0.41$. Its shape is similar to that of the binary chain with the same correlation length, but its magnitude is $15$ times larger ($G(0)=0.040$ for the binary chain), because the binary chain at this $\xi$ has much weaker nearest-neighbour correlations. The range of the correlations sets the shape of $G$, and their strength its magnitude.
 
Since every step costs roughly the same, the optimal schedule is almost uniform: $U=0.499$ and $\Lambda^2=0.498$, a gain of only $0.3\%$, while the cosine schedule has coefficient $0.598$, $20\%$ worse, see Fig.~\ref{fig:english}(b). Starting with small steps no longer pays off: since each step costs $\propto G\,\Delta p_n^2$, unequal steps are penalised unless $G$ varies correspondingly, and here it varies by only a factor $1.5$, see Fig.~\ref{fig:english}(a). All three schedules reach their large-$N$ coefficients already for $N\approx10$. The reason is that the assumptions underlying the small-step expression~\eqref{eq:Bcont} hold even for large steps. First, Eq.~\eqref{eq:Bn} neglects screening, i.e., the factor $(1-q_n)^{d-1}$, the probability that no site between a pair at distance $d$ is co-revealed. Here $I(1)=0.53$ and $I(2)=0.06$, so the information is carried almost entirely by adjacent pairs, $d=1$, for which this factor is exactly one. Second, replacing the sum~\eqref{Bgeo} by the integral~\eqref{eq:Bcont} requires $G$ to change little over a step, which holds for the nearly flat $G$ of Fig.~\ref{fig:english}(a) even for $\Delta p\sim0.1$.

\section{Conclusion}

For Markov data, generation in an MDM is shown to be an exactly solvable RSA process: the gap statistics provide both the path error and the output statistics exactly, for any number of steps. For many steps the error reduces to a thermodynamic metric, and the schedules that minimize it are its geodesics. The exact solutions clarifies in particular the interplay between the correlations and output error and quality. The longer the correlations, the more the error concentrates at the start of generation. Schedules that start slowly then gain more, both in path error and in perplexity, and also need fewer steps to reach their asymptotic behaviour. For predictable data with long correlations, the perplexity of the output under the uniform schedule can grow even when the information lost per token saturates, because each spurious switch becomes more costly. The geodesic schedule, although constructed only to minimize the path error, reduces the perplexity as well. On the other hand, for short-range data like English, a uniform schedule is already close to optimal. 

The RSA picture also applies to rules for \emph{which} positions to reveal, beyond the schedule, which fixes only \emph{how many}. Any rule in which the selection probability of a site depends only on the nearest revealed tokens keeps Eq.~\eqref{eq:master} closed and corresponds to the acceptance function of interacting RSA~\cite{Evans1993,Baule2019}, so practical samplers of this kind~\cite{Chang2022,BenHamu2025,DAPD2026,Bansal2026} can be analysed within the same framework.

Natural language has power-law correlations~\cite{EbelingPoschel1994,Altmann2012,LinTegmark2017} and hierarchical structure~\cite{Chomsky1956,LinTegmark2017,Sainburg2019} that bigram data lack. Measured mutual information decays exponentially at short distances, as for Markov data, and crosses over to a power law at longer distances~\cite{Sainburg2019}. For such data revealed tokens no longer shield, and the exact solution does not apply. Extending the approach beyond Markov data is left for future work.

\begin{appendix}

\section{Relation between $B$ and the training loss}
\label{sec:mdmloss}

MDMs are trained by masking a data sequence at random and minimising the cross-entropy of the network's predictions $p_\theta(x_i|\vec x_S)$ for the masked tokens~\cite{Sahoo2024,Shi2024,Ou2025}. The connection with $B$ follows from the distribution of sequences generated by the parallel sampler, $Q_{\rm out}(\vec x)=\sum_{\bsig}Q(\vec x,\bsig)$, where in this paragraph $Q$ denotes Eq.~\eqref{Qprob} with $p_{\rm data}$ replaced by $p_\theta$. Since the reveal steps are drawn from $w(\bsig)$ independently of the tokens, Jensen's inequality gives the variational bound
\be
-\ln Q_{\rm out}(\vec x)&\le&\sum_{\bsig}w(\bsig)\Big[-\ln\frac{Q(\vec x,\bsig)}{w(\bsig)}\Big]\nonumber\\
&=&\sum_{\bsig}w(\bsig)\sum_{n=0}^{N-1}\sum_{i\in U_n}-\ln p_\theta(x_i|\vec x_{S_n})\nonumber\\
\label{eq:elbo}
\ee
The training loss is the average of the right-hand side over the data,
\be
J_\theta=\sum_{n=0}^{N-1}\Big\langle\sum_{i\in U_n}-\ln p_\theta(X_i|\vec X_{S_n})\Big\rangle
\label{eq:loss}
\ee
a sum of cross-entropies, one for each masked token, each of which is minimized by the true conditional. The minimizer is therefore the ideal denoiser, for any schedule and any $N$. At the minimum, $Q$ is the parallel sampler of Eq.~\eqref{Qprob}, and since $P(\vec x,\bsig)/w(\bsig)=p_{\rm data}(\vec x)$,
\be
\min_\theta J_\theta&=&-\left\langle\ln\frac{Q(\vec X,\bsig)}{w(\bsig)}\right\rangle\nonumber\\
&=&-\left\langle\ln\frac{P(\vec X,\bsig)}{w(\bsig)}\right\rangle+\left\langle\ln\frac{P(\vec X,\bsig)}{Q(\vec X,\bsig)}\right\rangle\nonumber\\
&=&H(p_{\rm data})+B
\label{eq:lossB}
\ee
where $H(p_{\rm data})=-\langle\ln p_{\rm data}(\vec X)\rangle$ is the entropy of the data. Even a perfectly trained network thus has a loss that exceeds the data entropy by $B$: the path error is the part of the variational bound that cannot be removed by training, only by the choice of $N$ and the schedule. In practice MDMs are trained with the continuous-time limit of Eq.~\eqref{eq:loss}, $N\to\infty$,
\be
J_\theta=\int_0^1\frac{dp}{1-p}\,\Big\langle\sum_{i\notin S}-\ln p_\theta(X_i|\vec X_{S})\Big\rangle_{S\sim p}
\label{eq:losscont}
\ee
where each site belongs to $S$ independently with probability $p$, as follows from Eq.~\eqref{eq:loss} with $\P(i\in U_n|i\notin S_n)=q_n=\Delta p_n/(1-p_n)$. In this limit $B\to0$ and the minimal loss equals $H(p_{\rm data})$~\cite{Ou2025}. The training objective therefore measures the quality of the denoiser, whereas $B$ measures the additional error incurred when generation uses finitely many steps.

\section{Numerical validation and details}\label{app:numerics}

The master equation~\eqref{eq:master} and the path error~\eqref{eq:Bexact} are
exact. We checked them against two computations that do not use the gap
representation: exact enumeration of all generation histories for short sequences
(App.~\ref{app:enum}), and direct simulation of the parallel sampler for long
sequences (App.~\ref{app:mc}). We use the binary chain of Eq.~\eqref{eq:Ising},
parametrised by $\lambda=e^{-1/\xi}$ and $f=(1-\lambda)/2$, and a random asymmetric
three-state chain, generated with a fixed seed, to check that nothing relies on the
symmetry of the binary chain. The following schedules are used in the numerical validation.

\subsection{Masking schedules}\label{app:schedule}

A schedule fixes the revealed fraction $p_n$ after $n$ of the $N$ steps; in step $n$
an expected number $L\Delta p_n$ of tokens is revealed, i.e., a fraction
$q_n=\Delta p_n/(1-p_n)$ of the tokens still masked. We compare three schedules,
written in terms of the reduced time $\tau=n/N$.

(i)~The \emph{uniform} schedule, $p_n=n/N$, reveals the same expected number of
tokens, $L/N$, in every step. Since fewer tokens remain masked as generation
proceeds, it reveals an increasing fraction of them, $q_n=1/(N-n)$, ending with
$q_{N-1}=1$.

(ii)~The \emph{cosine} schedule, $p_n=1-\cos(\pi n/2N)$, is the schedule of
MaskGIT~\cite{Chang2022}, where the masked fraction $1-p=\cos(\pi\tau/2)$ decreases
slowly at first and quickly at the end. It reveals few tokens in the first steps,
$\Delta p_0=1-\cos(\pi/2N)\simeq\pi^2/(8N^2)$, compared with $1/N$ for the uniform
schedule, and more in the last ones, $\Delta p_{N-1}=\sin(\pi/2N)\simeq\pi/(2N)$. It
therefore takes small steps at the start of generation and large steps at the end.

The continuum forms of these schedules and the \emph{optimal} schedule are discussed in Sec.~\ref{sec:results}; the optimal schedule is derived in Sec.~\ref{sec:geo}.

\subsection{Exact enumeration}\label{app:enum}

\begin{table}
\caption{Exact enumeration of all generation histories, at the largest $N$
enumerated. $\delta\mathcal P$ is the largest deviation of the pair counts of the
master equation from enumeration, and $\delta B$ the relative deviation of
Eq.~\eqref{eq:Bexact} from the enumerated path relative entropy $B$.}
\label{tab:enum}
\begin{ruledtabular}
\begin{tabular}{lcccccc}
chain & $L$ & schedule & $N$ & $\delta\mathcal P$ & $\delta B$ & $B$\\\hline
binary, $\xi=2$ & 8 & uniform & 4 & $4\times10^{-15}$ & $5\times10^{-16}$ & 0.3068\\
binary, $\xi=2$ & 8 & cosine & 4 & $2\times10^{-15}$ & $6\times10^{-16}$ & 0.3442\\
binary, $\xi=5$ & 8 & uniform & 4 & $8\times10^{-15}$ & $1\times10^{-15}$ & 0.5501\\
binary, $\xi=5$ & 8 & cosine & 4 & $1\times10^{-14}$ & $2\times10^{-15}$ & 0.5405\\
binary, $\xi=20$ & 8 & uniform & 4 & $2\times10^{-14}$ & $<10^{-16}$ & 0.7634\\
binary, $\xi=20$ & 8 & cosine & 4 & $2\times10^{-14}$ & $2\times10^{-16}$ & 0.6063\\
three-state & 6 & uniform & 3 & $1\times10^{-15}$ & $5\times10^{-16}$ & 0.1028\\
three-state & 6 & cosine & 3 & $4\times10^{-16}$ & $1\times10^{-16}$ & 0.1227\\
\end{tabular}
\end{ruledtabular}
\end{table}

For short sequences every generation history can be listed. A history is specified
by the reveal steps $\bsig=(\sigma_1,\dots,\sigma_L)$, which occur with probability
$w(\bsig)=\prod_i\Delta p_{\sigma_i}$, and by the tokens $\vec x$. For each of the
$N^L$ step assignments and $V^L$ sequences we compute the path probability of the
parallel sampler, Eq.~\eqref{Qprob}. It requires the conditional
$p_{\rm data}(x_i|\vec x_{S_n})$ of every token revealed in step $n$, given the
tokens revealed before. In the enumeration this conditional is obtained by brute
force from the definition of a conditional probability,
\be
p_{\rm data}(x_i|\vec x_S)&=&\frac{p_{\rm data}(x_i,\vec x_S)}{p_{\rm data}(\vec x_S)}\\
p_{\rm data}(\vec x_S)&=&\sum_{\{x_j:\,j\notin S\}}p_{\rm data}(\vec x),
\ee
where the marginals are computed by summing the full sequence probability,
Eq.~\eqref{pdatamarkov}, over all values of the tokens that are not fixed. For
$L\le8$ these sums have at most $V^L$ terms and are evaluated explicitly. This
computation uses neither shielding nor the bridge formulas~\eqref{eq:marg},%
\eqref{eq:marg2} nor the gap densities. Summing the path probabilities over
histories gives the expected pair counts $\mathcal P_{bd}$ in the output, which the
master equation predicts, and the path relative entropy~\eqref{eq:B}, which
Eq.~\eqref{eq:Bexact} predicts. Separately, the distributions~\eqref{eq:marg},\eqref{eq:marg2}, including the conventions at the
ends of the sequence, were checked against the same brute-force conditionals for
$L\le10$.

Table~\ref{tab:enum} lists the comparison at the largest number of steps enumerated.
In all enumerated cases, $N=1,\dots,4$ for the binary chains and $N=1,\dots,3$ for
the three-state chain, the master equation and Eq.~\eqref{eq:Bexact} agree with
enumeration to machine precision.

\subsection{Direct simulation}\label{app:mc}

For long sequences we simulate the parallel sampler. In step $n$, every masked site
is selected with probability $q_n$, its nearest revealed sites are determined from
the state \emph{before} the step, and its token is drawn from the
conditional~\eqref{eq:marg}; all selected sites are then updated simultaneously. From $10^4$ samples of length $L=1024$ we estimate $\mathcal P_{bd}$, the
first-token distribution $\phi_d$ and $\Delta\CE$, Eq.~\eqref{eq:CE}, and compare
them with the master equation. As a control
we simulate the exact sampler, which must give $\Delta\CE=0$. Table~\ref{tab:mc} lists the comparison. All deviations, of $\Delta\CE$ and of
every pair count, are below two standard errors.

\begin{widetext}

\begin{table}[h]
\caption{Direct simulation of the parallel sampler ($L=1024$, $10^4$ samples per
row) against the master equation: excess cross-entropy per token, $\Delta\CE/L$,
with the standard error of the simulation. The last row simulates the exact
sampler, for which $\Delta\CE=0$.}
\label{tab:mc}
\begin{ruledtabular}
\begin{tabular}{lcccc}
chain & schedule & $N$ & master equation & simulation\\\hline
binary, $\xi=2$ & uniform & 8 & 0.05137 & $0.05147\pm0.00019$\\
binary, $\xi=10$ & uniform & 4 & 0.34969 & $0.35036\pm0.00040$\\
binary, $\xi=10$ & uniform & 16 & 0.07754 & $0.07735\pm0.00026$\\
binary, $\xi=10$ & cosine & 64 & 0.01261 & $0.01280\pm0.00022$\\
binary, $\xi=50$ & uniform & 16 & 0.14928 & $0.14978\pm0.00034$\\
three-state & cosine & 8 & 0.02022 & $0.02024\pm0.00019$\\
exact sampler, $\xi=10$ & -- & -- & 0 & $-0.00027\pm0.00020$\\
\end{tabular}
\end{ruledtabular}
\end{table}
\end{widetext}

\section{Derivation of $G(p_n)$ Eq.~(\ref{Gexp2})}\label{app:Gexp2}

Eq.~\eqref{Gexp} contains the interior gaps $(a,\ell,c)$, with $a,c$ tokens of the
vocabulary; gaps with $\ell<2$ contain no pair and do not contribute. The
dependence on the boundary tokens enters only through the combination
\begin{equation}
\bar I_{kk'}(\ell)\equiv\sum_{a,c}\pi_a(T^{\ell+1})_{ac}\,I^{(a,\ell,c)}_{kk'}
\label{eq:Ibar_def}
\end{equation}
so that
\begin{equation}
G(p)=\frac{p^2}{(1-p)^2}\sum_{\ell\ge2}(1-p)^\ell\sum_{1\le k<k'\le\ell}\bar I_{kk'}(\ell)
\label{eq:G_Ibar}
\end{equation}
We evaluate $\bar I_{kk'}(\ell)$ in four steps. Throughout, the boundary tokens sit at
sites $0$ and $\ell+1$ of the gap, and the pair at sites $k<k'$ carries the
tokens $v$ and $u$.

\paragraph*{Step 1: the ratio in Eq.~(\ref{eq:Igap}).}
With Eqs.~\eqref{eq:marg} and~\eqref{eq:marg2},
\be
m^{(a,\ell,c)}_{k,k'}(v,u)&=&\frac{(T^{k})_{av}(T^{k'-k})_{vu}(T^{\ell+1-k'})_{uc}}{(T^{\ell+1})_{ac}}\\
m^{(a,\ell,c)}_k(v)&=&\frac{(T^{k})_{av}(T^{\ell+1-k})_{vc}}{(T^{\ell+1})_{ac}}\\
m^{(a,\ell,c)}_{k'}(u)&=&\frac{(T^{k'})_{au}(T^{\ell+1-k'})_{uc}}{(T^{\ell+1})_{ac}}
\ee
In the ratio of the joint distribution to the product of the marginals, the factors
$(T^k)_{av}$ and $(T^{\ell+1-k'})_{uc}$ cancel, and one factor $(T^{\ell+1})_{ac}$ remains:
\begin{equation}
\frac{m^{(a,\ell,c)}_{k,k'}(v,u)}{m^{(a,\ell,c)}_k(v)\,m^{(a,\ell,c)}_{k'}(u)}
=\frac{(T^{k'-k})_{vu}\,(T^{\ell+1})_{ac}}{(T^{k'})_{au}\,(T^{\ell+1-k})_{vc}}
\label{eq:ratio}
\end{equation}

\paragraph*{Step 2: the weight is the data distribution.}
Inserting Eq.~\eqref{eq:Igap} into Eq.~\eqref{eq:Ibar_def}, the boundary weight
multiplies the joint distribution of the pair, and the denominator $(T^{\ell+1})_{ac}$ of
Eq.~\eqref{eq:marg2} cancels:
\be
&&\pi_a(T^{\ell+1})_{ac}\,m^{(a,\ell,c)}_{k,k'}(v,u)\nonumber\\
&&=\pi_a(T^{k})_{av}(T^{k'-k})_{vu}(T^{\ell+1-k'})_{uc}
\ee
By Eqs.~\eqref{pdatamarkov} and~\eqref{matrixnorm}, the right-hand side is the
probability that four tokens of the data at the positions $0,k,k',\ell+1$ take the values
$a,v,u,c$. Hence $\bar I_{kk'}(\ell)$ is the average of the logarithm of
Eq.~\eqref{eq:ratio} over the data:
\be
\bar I_{kk'}(\ell)&=&\Big\langle\ln(T^{k'-k})_{vu}+\ln(T^{\ell+1})_{ac}\nonumber\\
&&-\ln(T^{k'})_{au}-\ln(T^{\ell+1-k})_{vc}\Big\rangle
\label{eq:Ibar_logs}
\ee
This step requires the average over the boundary tokens: for fixed $a,c$, the tokens
inside the gap are not distributed as in the data.

\paragraph*{Step 3: each logarithm is a mutual information.}
Each term of Eq.~\eqref{eq:Ibar_logs} involves two of the four tokens, a distance
$m$ apart. By stationarity their joint probability is $\pi_x(T^m)_{xy}$, so that
\begin{equation}
\big\langle\ln(T^m)_{xy}\big\rangle=\sum_{x,y}\pi_x(T^m)_{xy}\ln(T^m)_{xy}
=I(m)-H_1
\end{equation}
where $H_1=-\sum_v\pi_v\ln\pi_v$ is the entropy of a single token. The second
equality follows from the definition of $I(m)$ and
$\sum_x\pi_x(T^m)_{xy}=\pi_y$:
\begin{equation}
I(m)=\sum_{x,y}\pi_x(T^m)_{xy}\ln(T^m)_{xy}-\sum_y\pi_y\ln\pi_y 
\end{equation}
Terms with $(T^m)_{xy}=0$ carry zero weight and do not contribute.

\paragraph*{Step 4: the entropies cancel.}
The four logarithms in Eq.~\eqref{eq:Ibar_logs} refer to the distances $k'-k$
(the pair), $\ell+1$ (the two boundaries), $k'$ (left boundary to the right site
of the pair), and $\ell+1-k$ (left site of the pair to the right boundary). Two
enter with a plus and two with a minus sign, so the single-token entropies $H_1$
cancel, and
\begin{equation}
\bar I_{kk'}(\ell)=I(k'-k)+I(\ell+1)-I(k')-I(\ell+1-k)
\label{eq:Ibar_result}
\end{equation}

\paragraph*{Pair variables.}
We label a pair in a gap by $l=k$, $d=k'-k$ and $r=\ell+1-k'$: the distances from
the left boundary to the first site of the pair, between the two sites, and from the
second site to the right boundary. The constraints $\ell\ge2$ and $1\le k<k'\le\ell$
are equivalent to $l,d,r\ge1$, and each triple $(l,d,r)$ corresponds to exactly one
gap length $\ell=l+d+r-1$ and one pair $(k,k')$. In these variables the distances
in Eq.~\eqref{eq:Ibar_result} are $k'-k=d$, $\ell+1=d+l+r$, $k'=d+l$ and
$\ell+1-k=d+r$, so that $\bar I_{kk'}(\ell)=\bar I(l,d,r)$, and the prefactor in
Eq.~\eqref{eq:G_Ibar} becomes
\begin{equation}
\frac{p^2}{(1-p)^2}(1-p)^\ell=p^2(1-p)^{l+d+r-3}
\end{equation}
Replacing the sums over $\ell$, $k$ and $k'$ in Eq.~\eqref{eq:G_Ibar} by a sum over
$l,d,r\ge1$ gives Eq.~\eqref{Gexp2}.

\section{Derivation of $U$, Eq.~(\ref{Uex})}\label{app:Uex}

The large-$N$ coefficient of the uniform schedule is $U=\int_0^1G(p)\,dp$, which we
evaluate from Eq.~\eqref{Gexp2}. We write $s=1-p$ and introduce the generating function
of the mutual information
\begin{equation}
\psi(s)=\sum_{m\ge1}I(m)\,s^{m-1}
\label{eq:psi}
\end{equation}
which converges for $0\le s\le1$ since $I(m)$ decays exponentially beyond the
correlation length. Its derivatives are
\be
\psi'(s)&=&\sum_{m\ge2}(m-1)\,I(m)\,s^{m-2}\\
\psi''(s)&=&\sum_{m\ge3}(m-1)(m-2)\,I(m)\,s^{m-3}
\ee
so that $\psi(0)=I(1)$ and $\psi'(0)=I(2)$.

\paragraph*{Step 1: summing over the outer distances.}
We insert Eq.~\eqref{Ibar} into Eq.~\eqref{Gexp2} and
treat the four terms separately, using $\sum_{l\ge1}s^l=s/p$.

(i)~In the term $I(d)$ the sums over $l$ and $r$ are geometric,
$\sum_{l,r\ge1}p^2s^{l+d+r-3}=s^{d-1}$, so that it contributes
$\sum_{d\ge1}s^{d-1}I(d)=\psi(s)$.

(ii)~In the term $-I(d+l)$ the sum over $r$ gives $p\,s^{l+d-2}$. With $m=d+l$, each
$m\ge2$ arises from $m-1$ pairs $(d,l)$, so that the term contributes
\be
-p\sum_{m\ge2}(m-1)s^{m-2}I(m)=-p\,\psi'(s)
\ee
By the symmetry $l\leftrightarrow r$
the term $-I(d+r)$ gives the same.

(iii)~In the term $I(d+l+r)$ we set $m=d+l+r$. Each $m\ge3$ arises from
$\binom{m-1}{2}=\frac12(m-1)(m-2)$ triples $(l,d,r)$, so that the term contributes
\be
\frac{p^2}{2}\sum_{m\ge3}(m-1)(m-2)s^{m-3}I(m)=\frac{p^2}{2}\psi''(s)
\ee

Collecting the four terms,
\begin{equation}
G(p)=\psi(s)-2p\,\psi'(s)+\frac{p^2}{2}\,\psi''(s),\qquad s=1-p
\label{eq:Gpsi}
\end{equation}
As a check, at $p=0$ Eq.~\eqref{eq:Gpsi} gives $G(0)=\psi(1)=\sum_{m\ge1}I(m)$, the total
pair information, which sets the metric at the start of generation (see Sec.~\ref{sec:results}).

\paragraph*{Step 2: integrating over $p$.}
Changing the integration variable to $s=1-p$,
\begin{equation}
U=\int_0^1\Big[\psi(s)-2(1-s)\,\psi'(s)+\frac{(1-s)^2}{2}\,\psi''(s)\Big]ds.
\label{eq:Uint}
\end{equation}
Using integration by parts yields in the second term
\begin{equation}
-2\int_0^1(1-s)\,\psi'(s)\,ds=2\psi(0)-2\int_0^1\psi(s)\,ds,
\end{equation}
and in the term
\be
\frac12\int_0^1(1-s)^2\,\psi''(s)\,ds&=&-\frac12\psi'(0)+\int_0^1(1-s)\,\psi'(s)\,ds\nonumber\\
&=&-\frac12\psi'(0)-\psi(0)+\int_0^1\psi(s)\,ds\nonumber\\
\ee
The boundary terms at $s=1$ vanish because of the factors $(1-s)$ and $(1-s)^2$. Adding
the three contributions to Eq.~\eqref{eq:Uint}, the integrals of $\psi$ appear with
coefficients $1-2+1=0$ and cancel. Only the boundary terms at $s=0$ remain:
\begin{equation}
U=\psi(0)-\frac12\psi'(0)=I(1)-\frac12I(2),
\end{equation}
which is Eq.~\eqref{Uex}. Correlations at all distances enter $G(p)$ through $\psi$,
but after integration only the boundary values $\psi(0)=I(1)$ and $\psi'(0)=I(2)$
survive.

\end{appendix}

\end{document}